\documentclass[aps, prl, twocolumn, nofootinbib, floatfix, superscriptaddress]{revtex4-1}

\usepackage{comment}
\usepackage{graphicx}
\usepackage{dcolumn}
\usepackage{bm}
\usepackage{subfigure}
\usepackage{epsfig}
\usepackage{amsmath}
\usepackage{amsfonts}
\usepackage{graphicx}
\usepackage{amssymb}
\usepackage{nccmath}
\usepackage{amssymb,amsmath,amsfonts}
\usepackage{xfrac}
\usepackage{color}
\usepackage{tikz}
\usepackage[T1]{fontenc}
\usepackage[utf8]{inputenc}
\usepackage{ulem}

\usepackage{tcolorbox}
\usepackage{hyperref}
\usepackage{booktabs}
\hypersetup{colorlinks=true, citecolor=red, linkcolor=blue, urlcolor=blue}

\definecolor{rossos}{cmyk}{0,1,1,0.55}
\definecolor{bluscuro}{rgb}{0.15, 0.2, .85}
\definecolor{bluchiaro}{cmyk}{1,.3,0.,0.1}

\let\oldsqrt\sqrt
\def\sqrt{\mathpalette\DHLhksqrt}
\def\DHLhksqrt#1#2{%
\setbox0=\hbox{$#1\oldsqrt{#2\,}$}\dimen0=\ht0
\advance\dimen0-0.2\ht0
\setbox2=\hbox{\vrule height\ht0 depth -\dimen0}%
{\box0\lower0.4pt\box2}}

\newcommand{\sss}[1]{{\scriptscriptstyle{#1}}}
\newcommand{\boldmathsymbol}[1]{{\ensuremath{\boldsymbol{#1}}}}

\newcommand{\uPl}{\mathrm{Pl}}

\newcommand{\usssPl}{\sss{\uPl}}

\newcommand{\Mp}{M_\usssPl}

\newcommand{\beq}{\begin{equation}}
\newcommand{\eeq}{\end{equation}}
\newcommand{\bea}{\begin{equation}\begin{aligned}}
\newcommand{\eea}{\end{aligned}\end{equation}}

\newlength{\wsingfig}
\newlength{\wdblefig}
\newlength{\wquadfig}
\newlength{\wtriplefig}
\newcommand{\Eq}[1]{Eq.~(\ref{#1})}

\newcommand{\Fig}[1]{Fig.~{\ref{#1}}}

\newcommand{\be}{\begin{equation}}

\begin{document}
\rightline{YITP-24-22}
\title{New probe of non-Gaussianities with primordial black hole induced gravitational waves}

\author{Theodoros Papanikolaou}
\email{t.papanikolaou@ssmeridionale.it}
\affiliation{Scuola Superiore Meridionale, Largo San Marcellino 10, 80138 Napoli, Italy}
\affiliation{Istituto Nazionale di Fisica Nucleare (INFN), Sezione di Napoli, Via Cinthia 21, 80126 Napoli, Italy}
\affiliation{National Observatory of Athens, Lofos Nymfon, 11852 Athens, 
Greece}

\author{Xin-Chen He}
\email{xinchenhe@mail.ustc.edu.cn}
\affiliation{Department of Astronomy, School of Physical Sciences, University of Science and Technology of China, Hefei 230026, China}
\affiliation{CAS Key Laboratory for Researches in Galaxies and Cosmology, School of Astronomy and Space Science, University of Science and Technology of China, Hefei, Anhui 230026, China}
\affiliation{Deep Space Exploration Laboratory, Hefei 230088, China}

\author{Xiao-Han Ma}
\email{mxh171554@mail.ustc.edu.cn}
\affiliation{Department of Astronomy, School of Physical Sciences, University of Science and Technology of China, Hefei 230026, China}
\affiliation{CAS Key Laboratory for Researches in Galaxies and Cosmology, School of Astronomy and Space Science, University of Science and Technology of China, Hefei, Anhui 230026, China}
\affiliation{Deep Space Exploration Laboratory, Hefei 230088, China}
\affiliation{Kavli Institute for the Physics and Mathematics of the Universe (WPI), UTIAS The University of Tokyo, Kashiwa, Chiba 277-8583, Japan}

\author{Yi-Fu Cai}
\email{yifucai@ustc.edu.cn}
\affiliation{Department of Astronomy, School of Physical Sciences, University of Science and Technology of China, Hefei 230026, China}
\affiliation{CAS Key Laboratory for Researches in Galaxies and Cosmology, School of Astronomy and Space Science, University of Science and Technology of China, Hefei, Anhui 230026, China}
\affiliation{Deep Space Exploration Laboratory, Hefei 230088, China}

\author{Emmanuel N. Saridakis}
\email{msaridak@noa.gr}
\affiliation{National Observatory of Athens, Lofos Nymfon, 11852 Athens, 
Greece}
\affiliation{Department of Astronomy, School of Physical Sciences, University of Science and Technology of China, Hefei 230026, China}
\affiliation{CAS Key Laboratory for Researches in Galaxies and Cosmology, School of Astronomy and Space Science, University of Science and Technology of China, Hefei, Anhui 230026, China}
\affiliation{Departamento de Matem\'{a}ticas, Universidad Cat\'{o}lica del Norte, Avda. Angamos 0610, Casilla 1280 Antofagasta, Chile}

\author{Misao Sasaki}
\email{misao.sasaki@ipmu.jp}
\affiliation{Kavli Institute for the Physics and Mathematics of the Universe (WPI), UTIAS The University of Tokyo, Kashiwa, Chiba 277-8583, Japan}
\affiliation{Center for Gravitational Physics, Yukawa Institute for Theoretical Physics, Kyoto University, Kyoto 606-8502, Japan}
\affiliation{Leung Center for Cosmology and Particle Astrophysics,
National Taiwan University, Taipei 10617}



\begin{abstract}
We propose a new probe of primordial non-Gaussianities (NGs) through the observation of gravitational waves (GWs) induced by ultra-light ($M_{\text{PBH}}< 10^{9}\rm{g}$) primordial black holes (PBHs). Interestingly enough, the existence of primordial NG can leave imprints on the clustering properties of PBHs and the spectral shape of induced GW signals. Focusing on a scale-dependent local-type NG, we identify a distinct double-peaked GW energy spectrum that, contingent upon $M_{\text{PBH}}$ and the abundance of PBHs at the time of formation, denoted as $\Omega_\mathrm{PBH,f}$, may fall into the frequency bands of upcoming GW observatories, including LISA, ET, SKA, and BBO. Thus, such a signal can serve as a novel portal for probing primordial NGs. Intriguingly, combining BBN bounds on the GW amplitude, we find for the first time the joint limit on the product of the effective non-linearity parameter for the primordial tri-spectrum, denoted by $\bar{\tau}_\mathrm{NL}$, and the primordial curvature perturbation power spectrum $\mathcal{P}_{\cal R}(k)$, which reads as $\bar{\tau}_\mathrm{NL} \mathcal{P}_{\cal R}(k) < 4\times 10^{-20} \Omega^{-17/9}_\mathrm{PBH,f} \left( \frac{M_{\rm PBH}}{10^4\mathrm{g}} \right)^{-17/9}$.
\end{abstract}


\maketitle


{\it{Introduction}} -- 
Primordial non-Gaussianity (NG), as a fundamental statistical characteristic of primordial perturbations, arises from potential nonlinear processes during inflation~\cite{Byrnes:2010ft,Riotto:2010nh, Martin:2012pe, Namjoo:2012aa}. The investigation of primordial NG yields crucial insights into the early Universe, elucidating inflationary dynamics and field contents across a wide range of energy scales, complementing the search for primordial gravitational waves (GWs)~\cite{Cai:2018dig, Adshead:2021hnm, Domenech:2021ztg, Li:2023qua,Choudhury:2023fwk,Li:2024zwx}. As indicators of NG, high-order correlations of primordial perturbations have been well studied theoretically~\cite{Bartolo:2004if, Dalal:2007cu, Chen:2010xka}. However, 
despite the cosmic microwave background (CMB) measurements constraining primordial bispectra (e.g. $f^{\rm local}_{\rm NL}=-0.9\pm 5.1$~\cite{Planck:2019kim}), the observational limits on other higher-order correlations (namely trispectra) remain loose. 

Facing the limitations in observing high-order correlations, primordial black holes (PBHs)~\cite{1967SvA....10..602Z, Carr:1974nx, Carr:1975qj}, originating from large amplitude curvature perturbations during inflation (see \cite{Sasaki:2018dmp, LISACosmologyWorkingGroup:2023njw} for recent reviews), may provide a loophole in measuring primordial NGs on small scales. Currently, most efforts are made to constrain the PBH abundances across a broad range of mass scales~\cite{Carr:2020gox}. Meanwhile, the distribution of PBHs has recently drawn much attention because Gaussian primordial fluctuations cannot lead to clustering beyond a Poisson distribution~\cite{Ali-Haimoud:2018dau, Desjacques:2018wuu, MoradinezhadDizgah:2019wjf}. This implies that a possible detection of nontrivial spatial distributions of PBHs can strongly hint the existence of primordial NGs~\cite{Byrnes:2012yx, Young:2013oia, Pattison:2017mbe, Franciolini:2018vbk, Suyama:2019cst, Ezquiaga:2019ftu, Figueroa:2020jkf, Kitajima:2021fpq, Auclair:2024jwj, Animali:2024jiz}.

Notably, ultra-light PBHs with $M_{\rm PBH} < 10^{9}\mathrm{g}$, which evaporate before Big Bang Nucleosynthesis (BBN) and hence escape from the current direct observational constraints on their abundance, can be produced naturally in the early Universe in many inflationary frameworks and beyond, such as preheating/reheating~\cite{Martin:2019nuw}, quantum diffusion~\cite{Pattison:2017mbe,Ezquiaga:2019ftu,Briaud:2023eae}, stochastic tunnelling~\cite{Animali:2022otk}, as well as within bouncing~\cite{Banerjee:2022xft} and quantum  cosmological~\cite{Papanikolaou:2023crz} frameworks acting as a probe of the thermal history of the Universe~\cite{Domenech:2020ssp,Domenech:2021wkk} and new physics phenomenology~\cite{Banerjee:2022xft,Papanikolaou:2023crz,Papanikolaou:2023oxq,Basilakos:2023xof, Barrow:1990he, Bhaumik:2022pil}. More specifically, a population of such light PBHs being modeled as a gravitationally interacting gas on large scales~\cite{Papanikolaou:2020qtd, Domenech:2020ssp} and possessing its own density perturbations, can give rise to early matter-dominated (eMD) phases before BBN~\cite{Garcia-Bellido:1996mdl, Hidalgo:2011fj, Suyama:2014vga, Zagorac:2019ekv}.

Interestingly enough, during the transition from the eMD era to the late radiation-dominated (lRD) era~\cite{Inomata:2019ivs, Inomata:2020lmk,Atal:2021jyo,Domenech:2021and,LISACosmologyWorkingGroup:2023njw,Papanikolaou:2022chm}, the aforementioned PBH density perturbations can evolve into large, rapidly oscillating inhomogeneities, resulting in observable and distinctive induced GWs~\cite{Papanikolaou:2020qtd, Domenech:2020ssp,Domenech:2023jve}. If there were primordial NGs, these GWs would be influenced and thus retain their information appropriately. In this Letter, we investigate the effects of a generalized scale-dependent local-type trispectrum on these GW signals, which, as will see below, can act as a novel probe to constrain primordial non-Gaussianities on very small scales, otherwise unconstrained by other observational probes.



{\it{The Gaussian PBH gas}}\label{sec:PBH_matter_PS_Gaussian} --
We begin by considering a population (``gas'') of PBHs randomly distributed in space forming on small scales in a radiation-dominated, homogeneous Universe. For simplicity, we assume that PBHs form from enhanced primordial curvature perturbations which are characterised by a sharply peaked power spectrum, so that the PBH mass distribution is almost monochromatic. One then encounters the formation of PBHs with mass of the order of the Hubble horizon mass at the time of PBH formation,
\begin{align}
\begin{aligned}
    M_{\rm PBH}&=\frac{4\pi\gamma \Mp^2}{H_{\rm f}}\,,
\end{aligned}
\end{align}   
with $H_\mathrm{f}$ being the Hubble parameter at the time of PBH formation and $\gamma\simeq 0.2$ representing the fraction of the Hubble horizon mass collapsing to PBHs~\cite{Musco:2008hv}.

Note that the formation of PBHs is a rare event, hence the mean separation distance between two PBHs is much larger than the PBH characteristic scale, $k_{\rm f}^{-1}$, with $k_{\rm f}$ being defined as $k_{\rm f}\equiv a_{\rm f}H_{\rm f}$, where $a_{\text{f}}$ is the scale factor at PBH formation. 
These PBHs can be effectively considered as a pressureless fluid on scales much larger than the mean separation distance, which is naturally identified as an ultraviolet (UV) cut-off, denoted by $k_{\rm UV}^{-1}\equiv (\gamma/\Omega_{\rm PBH,f})^{1/3}k_{\rm f}^{-1}$, where $\Omega_\mathrm{PBH,f}$ is the energy density fraction of PBHs at the time of formation. 

In the case of Gaussian primordial curvature perturbations, their peaks in the high peak limit are uncorrelated, implying that PBHs are randomly distributed in space~\cite{Ali-Haimoud:2018dau}. Consequently, their statistics follow the Poisson distribution~\cite{Ali-Haimoud:2018dau,Desjacques:2018wuu}, resulting in the PBH matter power spectrum~\cite{Papanikolaou:2020qtd},
\begin{equation}\label{eq:ppoi}
    {\cal P}_{\delta_\mathrm{PBH}, \mathrm{G}}(k)=\frac{2}{3\pi}\left(\frac{k}{k_{\rm UV}}\right)^3\,,
\end{equation}
where the subscript ``G'' denotes the Gaussian nature of the primordial curvature perturbation. 
We note that as the scale decreases and approaches $k_{\rm UV}$, the PBH matter power spectrum approaches unity. This signals the transition into the non-linear regime, where the perturbation theory and the PBH gas description no longer holds.

{\it{The non-Gaussian PBH gas}}
\label{sec:PBH_matter_PS_non_Gaussian}--
We now introduce conventional local-type NGs, represented by the non-linearity parameters $f_{\rm NL}$, $g_{\rm NL}$ and $\tau_{\rm NL}$. Essentially, NGs induce deviations from the Poissonian behavior in the spatial distribution of PBHs, leading to initial PBH clustering~\cite{Tada:2015noa,Young:2015kda,Suyama:2019cst}. 
This clustering effect is mainly governed by the primordial trispectrum, or more precisely, the non-linearity parameter $\tau_{\rm NL}$~\cite{Suyama:2019cst}.
Thus, working within the high-peak regime, i.e. $\nu\gg 1$, one gets that
\begin{eqnarray}\label{eq:ppbhng}
\mathcal{P}_{\delta_\mathrm{PBH}}(k)\simeq\mathcal{P}_{\delta_\mathrm{PBH},\rm Poisson}(k)+\nu^4\bar{\tau}_{\rm NL} \mathcal{P}_{\mathcal{R}}(k)\,. 
\end{eqnarray}
Here, $\bar{\tau}_{\rm NL}$ is an effective $\tau_{\rm NL}$ parameter defined as
\begin{eqnarray}\label{eq:eff_tau_NL}
 &&\bar{\tau}_{\rm NL} \equiv \left(\frac{4}{9\sigma_R}\right)^4 \int \frac{\mathrm{d}^3p_1 \mathrm{d}^3p_2}{(2\pi)^6}\tau_\mathrm{NL}(p_1,p_2,p_1,p_2)\cr 
&&\qquad\qquad\quad\times W^2_\mathrm{local}(p_1)W^2_\mathrm{local}(p_2)P_{\mathcal{R}}(p_1)P_{\mathcal{R}}(p_2)\,, 
\end{eqnarray}
where the primordial curvature spectrum $P_\mathcal{R}(k) = 2\pi^2\mathcal{P}_\mathcal{R}(k)/k^3$ and
$W_{\rm local}(k)\equiv(kR)^2W_R(k)$ is a smoothing function with scale $R$, where $W_R(k)$  smooths the energy density field over scales inside the Hubble horizon~\cite{Ando:2018qdb,Young:2019osy}. 
We mention that \Eq{eq:ppbhng} is derived in the large-scale limit when $kR\ll 1$, which is consistent with the PBH fluid description described above, since we consider scales $k<k_{\rm UV}\ll k_\mathrm{f}$. 

A non-zero $f_{\rm NL}$ can also induce PBH clustering, resulting in a clustering term proportional to $f^2_\mathrm{NL}$~\cite{Tada:2015noa, Young:2015kda}. However, this contribution is inherently included in the $\tau_\mathrm{NL}$ term from \Eq{eq:ppbhng} due to the inequality $\tau_{\rm NL} \geq \frac{36}{25} f_{\text{NL}}^2$~\cite{Suyama:2007bg}. In single-field inflation scenarios, where this inequality is saturated, $f^2_\mathrm{NL}$ can be replaced by $\tau_{\rm NL}$. In other scenarios, the $f^2_\mathrm{NL}$ contribution may constitute only a small part of $\tau_{\rm NL}$.

For a sharply peaked primordial spectrum, such as the one studied here, we may identify the smoothing scale $R$ as $R\simeq k^{-1}_{\text{f}}$. 
The peak height $\nu$ is defined as $\nu \equiv \delta_{\rm cr}/\sigma_{R}$, with $\delta_{\rm cr} \simeq 0.41$ representing the critical energy density contrast for monochromatic PBH formation~\cite{Musco:2018rwt} and $\sigma_R$ denoting the smoothed standard deviation of the energy density contrast. Note here that for monochromatic PBH mass distributions $\nu$ can be approximated as $k$-independent being of the order of $10$ in order to obtain a sufficiently high peak collapsing to form a PBH~\cite{Suyama:2019cst}.

In \Eq{eq:ppbhng}, we generalised the results in~\cite{Suyama:2019cst} by considering scale-dependent non-linearity parameters. This extension is motivated by the amplification of  NGs at small scales due to possible nonlinear processes during late-time inflation~\cite{Riotto:2010nh, Byrnes:2010ft,Martin:2012pe,Namjoo:2012aa}. Nevertheless, we mention that  \Eq{eq:ppbhng} can be equally applied to the case where $\tau_{\rm NL}$ is scale-independent. 
Interestingly enough, the clustering term in \Eq{eq:ppbhng} is proportional to the primordial curvature spectrum, implying that any non-vanishing $\tau_{\rm NL}$-type NG will give rise to clustering of PBHs, significantly altering the scaling of the PBH matter spectrum and introducing distinguishable effects on large scales.
It is however noted that \Eq{eq:ppbhng} can apply only to the scales that has entered the horizon by the time of evaporation, i.e., $k>k_{\rm evap}$ where $k_{\rm evap}$ is the scale that enters the horizon at the time of evaporation. For the scales $k<k_{\rm evap}$, we expect a large suppression of ${\cal P}_{\delta_{\rm PBH}}$ due to causality.

{\it{The PBH gravitational potential}}\label{sec:PBH_Phi} --
The density inhomogeneities due to PBHs are isocurvature in nature, since there is no total energy density perturbations on superhorizon scales, i.e., at $k<k_{\rm UV}\ll k_{\rm f}$, at the time of PBH formation when $k_{\rm f}/a_{\rm f}=H_{\rm f}$~\cite{Inman:2019wvr}.
As the PBHs start to dominate the universe, the isocurvature perturbations convert to adiabatic perturbations, represented by the Bardeen potential $\Phi$. One then can straightforwardly show [See~\cite{Papanikolaou:2020qtd,Domenech:2020ssp} for more details] that the power spectrum for 
$\Phi$ right after PBH evaporation is directly related to the PBH matter power spectrum as
\beq\label{eq:P_Phi_PBH_full}
\mathcal{P}_{\Phi}(k)=S^2_\Phi(k)\left(5+\frac{8}{9}\frac{k^2}{{k}^2_\mathrm{d}}\right)^{-2}\mathcal{P}_{\delta_\mathrm{PBH}}(k)\,,
\eeq
where $\mathcal{P}_{\delta_\mathrm{PBH}}(k)$ is given by \Eq{eq:ppbhng} with $k_\mathrm{d}$ being the scale crossing the Hubble horizon at the onset of the PBH-dominated era, and $S_\Phi(k) \simeq \left(k/k_\mathrm{evap}\right)^{-1/3}$ represents the suppression factor due to the non-zero pressure of the radiation fluid~\cite{Inomata:2020lmk}. 
Interestingly, $\Phi$ sources GWs at second order as we will see below.

For quantitative analysis, we specify a concrete form of $\tau_{\rm NL}$ and the primordial curvature spectrum $\mathcal{P}_{\mathcal{R}}(k)$. 
For $\tau_\mathrm{NL}$ we propose a phenomenological parametrization with a log-normal distribution peaked at the PBH formation scale,
\begin{eqnarray}\label{eq:tau_NL_log_normal}
\begin{split}
\tau_\mathrm{NL}& (k_1,k_2,k_3,k_4)  = \\ & \frac{\tau_\mathrm{NL}(k_\mathrm{f})}{6}\left[e^{-\frac{1}{2\sigma_\tau^2}\left(\ln^2\frac{k_1}{k_\mathrm{f}}+\ln^2\frac{k_2}{k_\mathrm{f}}\right)}+\;5\;\mathrm{perms}\right]\,.
\end{split}
\end{eqnarray}
This form is motivated by the case of a scale-dependent $f_{\rm NL}$. 
Note that in the equilateral limit, where $k_1=k_2=k_3=k_4 =k_\mathrm{f}$,  one obtains $\tau_\mathrm{NL}$ evaluated at $k_\mathrm{f}$. For our numerical analyses,  we adopt $\sigma_\tau = 0.3$. As for the primordial curvature perturbation spectrum $\mathcal{P}_{\mathcal{R}}(k)$, we also assume a log-normal spectrum peaked at $k=k_{\rm f}$ with the same width used for $\tau$, $\sigma=0.3$, on top of a nearly scale-invariant spectrum $\mathcal{P}_{\mathcal{R}}(k)\simeq 10^{-9}$ indicated by the Planck CMB data~\cite{Planck:2018vyg}. The peak value is $\mathcal{P}_{\mathcal{R}}(k_{\rm f}) \simeq 10^{-2}$.

\begin{figure*}[t!]
\centering
\includegraphics[width=0.49\textwidth]{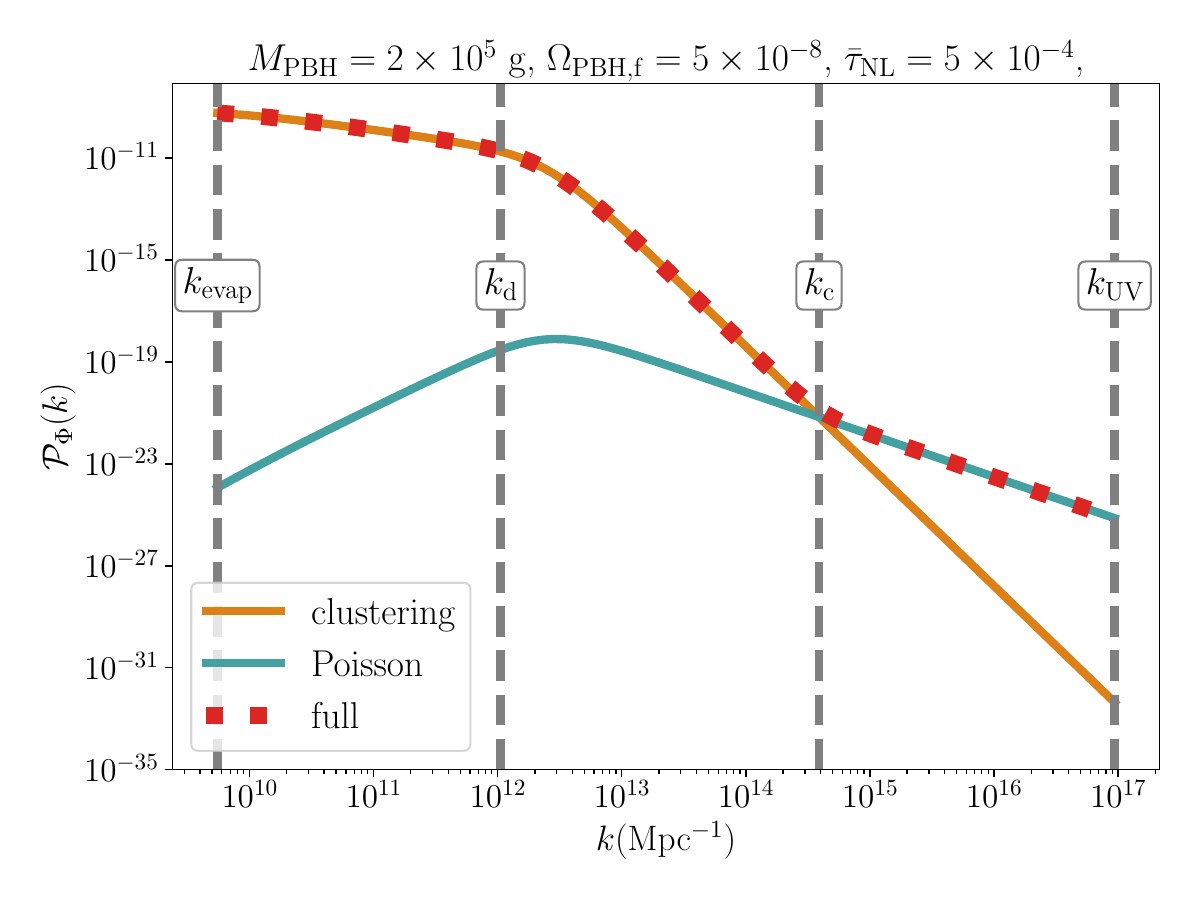}
\includegraphics[width=0.49\textwidth]{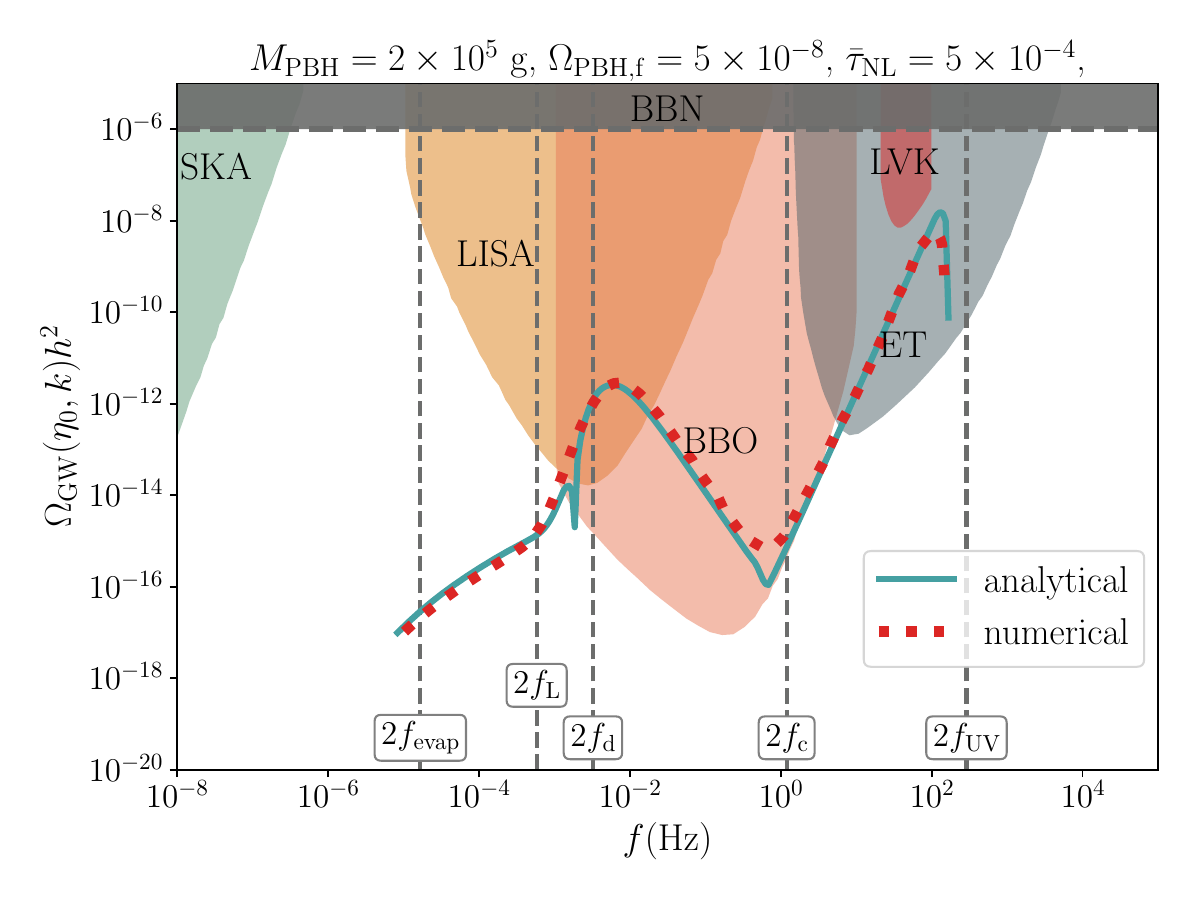}
\caption{\it{Left Panel: Power spectrum of the PBH gravitational potential for $M_\mathrm{PBH}= 2\times 10^{5}\mathrm{g}$ and $\Omega_\mathrm{PBH,f}=5\times 10^{-8}$ with $\bar{\tau}_{\rm NL} =  5 \times 10^{-4}$. The green solid curve depicts the Poisson term contribution from \Eq{eq:ppbhng} to $\mathcal{P}_{\Phi}(k)$, the orange solid one denotes the non-Gaussian modification through the clustering term in \Eq{eq:ppbhng}, and the dashed red line gives the total $\mathcal{P}_{\Phi}(k)$ in \Eq{eq:P_Phi_PBH_full}. 
Right Panel: The GW signal as a function of the GW frequency defined as $f\equiv \frac{k}{2\pi a_0}$ generated from PBH isocurvature induced adiabatic perturbations. The red dashed curve shows the exact numerical results and the green solid line gives our analytical approximation \eqref{eq:OmegaGWall}. We superimpose as well the sensitivity curves of GW experiments, including LIGO-VIRGO-KAGRA(LVK)\cite{KAGRA:2021kbb}, LISA~\cite{LISA:2017pwj, Karnesis:2022vdp}, ET~\cite{Maggiore:2019uih},SKA~\cite{Janssen:2014dka} and BBO~\cite{Harry:2006fi} for comparison. }}
\label{fig:P_Phi_Omega_GW}
\end{figure*}

In the following, we show the scalar power spectrum $\mathcal{P}_{\Phi}(k)$ by fixing the parameters of the problem at hand, namely the PBH abundance at formation time $\Omega_\mathrm{PBH,f}$, the PBH mass $M_\mathrm{PBH}$ and the effective $\bar{\tau}_\mathrm{NL}$. Thus, fixing $M_\mathrm{PBH} = 5 \times 10^{5} \mathrm{g}$, $\Omega_\mathrm{PBH,f} = 5 \times 10^{-8}$, and $\bar{\tau}_\mathrm{NL} =  10^{-3}$, corresponding to $\tau_\mathrm{NL}(k_\mathrm{f}) = 0.05$ within the log-normal $\tau_\mathrm{NL}$ parameterization \eqref{eq:tau_NL_log_normal}, the scalar power spectrum $\mathcal{P}_{\Phi}(k)$ is illustrated in the left panel of \Fig{fig:P_Phi_Omega_GW}.

As one may see from \Fig{fig:P_Phi_Omega_GW}, there exist four characteristic scales including $k_\mathrm{UV}$, $k_\mathrm{d}$ and $k_\mathrm{evap}$ mentioned above, and the critial scale $k_\mathrm{c}$ above which the PBH power spectrum is dominated by the Poisson term.  $k_\mathrm{UV}$, $k_\mathrm{d}$ and $k_\mathrm{evap}$ are explicitly related to $M_\mathrm{PBH}$ and $\Omega_\mathrm{PBH,f}$ as~\cite{Domenech:2020ssp}
\beq\label{eq:k_evap_d_UV}
\begin{split}
&\frac{k_\mathrm{UV}}{k_\mathrm{f}} = \left(\frac{\Omega_\mathrm{PBH,f}}{\gamma}\right)^{1/3}\quad,\quad\frac{k_\mathrm{d}}{k_\mathrm{f}} = \sqrt{2} \Omega_\mathrm{PBH,f}\,,
\\ & \frac{k_\mathrm{evap}}{k_\mathrm{f}} = \left(\frac{3.8g_{*}\Omega_\mathrm{PBH,f}}{960\gamma}\right)^{1/3}\left(\frac{M_\mathrm{PBH}}{\Mp}\right)^{-2/3}\,, 
\end{split}
\eeq
where $k_\mathrm{f}$ reads
\beq
\frac{k_\mathrm{f}}{10^{20} \mathrm{Mpc}^{-1}} \simeq \left(\frac{3.8g_{*}}{960\gamma^{3/4}}\right)^{2} \left(\frac{M_\mathrm{PBH}}{10^4\mathrm{g}}\right)^{-5/6}\left(\frac{\Omega_\mathrm{PBH,f}}{10^{-7}}\right)^{-1/3},
\eeq
and $g_{*}$ stands for the effective number of relativistic species at the epoch of PBH formation, being of the order of $100$ for PBHs evaporating before BBN~\cite{Carr:2020gox}. 

For $k_\mathrm{evap}$, $k_\mathrm{d}$, $k_\mathrm{UV}$, and $k_\mathrm{f}$, we have the hierarchy, $k_\mathrm{evap}<k_\mathrm{d}<k_\mathrm{UV}\ll k_\mathrm{f}$ which establishes two distinct regimes depending on the position of the critical scale $k_\mathrm{c}$. 
For extremely small values of $\bar{\tau}_\mathrm{NL}\mathcal{P}_{\cal R}$, we have $k_\mathrm{evap}<k_\mathrm{c}<k_\mathrm{d}$, which leads to no observable effect of NGs. Thus, the regime of interest for us is $k_\mathrm{d}<k_\mathrm{c}<k_\mathrm{UV}$, where a sizable effect of primordial NGs emerges at the level of the induced GW signal, as we discuss below.

{\it{Induced gravitational waves}}--
Let us now study the secondary GWs induced by scalar perturbations~\cite{Matarrese:1992rp,Matarrese:1993zf,Matarrese:1997ay,Mollerach:2003nq}  [See~\cite{Domenech:2021ztg} for a review].
Working in the Newtonian gauge, one gets the GW spectral abundance $\Omega_{\rm GW}$ which can be recast as~\cite{Ananda:2006af,Baumann:2007zm,Kohri:2018awv,Espinosa:2018eve}
\begin{eqnarray} \label{Tensor Power Spectrum}
&&
\!\!\!\!\!\!\!\!\!
\Omega_{\rm GW}(\eta,k) = \frac{2}{3}\int_{0}^{\infty} 
\mathrm{d}v\int_{|1-v|}^{1+v}\mathrm{d}u\! \left[ \frac{4v^2 - 
(1\!+\!v^2\!-\!u^2)^2}{4uv}\right]^{2}\nonumber\\
&&
\ \ \ \ \ \ \ \ \ \ \ \ \ \ \ \ \ \ \ \ \ \ \ \ \ \ \times
\overline{I^2}(x,u,v)\mathcal{P}_\Phi(kv)\mathcal{P}
_\Phi(ku)\,,
\end{eqnarray}
where $x\equiv k\eta$, $\eta$ is the conformal time and the bar denotes time averaging.
The two auxiliary variables $u$ and $v$ are defined respectively as $u \equiv 
|\boldmathsymbol{k} - \boldmathsymbol{q}|/k$ ,  $v \equiv q/k$.
The kernel function $I(x,u,v)$ contains information about the thermal history of the Universe ~\cite{Kohri:2018awv,Inomata:2019ivs,Inomata:2019zqy,Papanikolaou:2022chm} [See~\cite{Domenech:2021ztg} for its formal definition]. 

The kernel function $I(x,u,v)$ can be split in three pieces, representing contributions from three different epochs, as $I=I_\mathrm{eRD}+I_\mathrm{eMD}+I_\mathrm{lRD}$. 
Out of the above three terms, the last from the lRD era dominates due to the enhancement of $\Phi$ as a result of the sharp transition from eMD to lRD era caused by PBH evaporation~\cite{Inomata:2019ivs,Inomata:2020lmk}. Consequently, by replacing $\overline{I^2}$ with $\overline{I^2_{\rm lRD}}$ and ${\cal P}_\Phi$ with \Eq{eq:P_Phi_PBH_full}, the GW energy spectrum can be approximated as
${\Omega_{\rm GW}(k,\eta)}\simeq{\Omega_{\rm GW,lRD}(k,\eta)}$. 

Taking into account the cosmic evolution afterwards, the GW energy spectrum today, $\eta = \eta_0$, reads
\beq
\Omega_\mathrm{GW}(\eta_0,k)h^2 = 
c_{\rm g} \Omega_{\mathrm{r},0} h^2 \Omega_\mathrm{GW}(\eta_\mathrm{fGW},k)\,,
\eeq
where $c_\mathrm{g} = 0.4 \left(\frac{g_*(T_\mathrm{fGW})}{106.75}\right)^{-1/3}$,
$\Omega_{\mathrm{r},0}h^2 \simeq 4\times 10^{-5}$~\cite{Planck:2018vyg},
and $\Omega_\mathrm{GW}(\eta_\mathrm{fGW},k)$ is the GW spectral abundance evaluated at a time $\eta_\mathrm{fGW}$ during the $\mathrm{lRD}$ era by when the tensor modes can be treated as free GWs. 
For the modes of interest, one can safely assume that the tensor modes are propagating as free waves already after PBH evaporation.

As mentioned before, we focus on the most interesting part where $\bar{\tau}_\mathrm{NL}\mathcal{P}_\mathcal{R}$ is sizeable corresponding to the regime where $k_\mathrm{d}<k_\mathrm{c}<k_\mathrm{UV}$. In this case, the Poisson term primarily contributes to scales $2k_\mathrm{c}<k<2k_\mathrm{UV}$, while the clustering term dominates on larger scales $2k_\mathrm{evap}<k<2k_\mathrm{c}$.  

By splitting then appropriately the double integral~\Eq{Tensor Power Spectrum} into three regions, namely $[2k_\mathrm{evap},2k_\mathrm{d}]$, $[2k_\mathrm{d},2k_\mathrm{c}]$ and $[2k_\mathrm{c},2k_\mathrm{UV}]$,  and accounting for the resonant ($u+v\sim c^{-1}_\mathrm{s}$) and large $v,u$ ($u\sim v\gg 1$) contributions [See Appendix C of~\cite{Domenech:2020ssp}], one can derive after a lengthy but straightforward calculation an analytic approximate fitting formula for the GW energy spectrum in the different $k$ regions of interest, which can read in a compact form as
\begin{widetext}
\begin{align}\label{eq:OmegaGWall}
 \Omega_{\rm GW}(\eta_0, k)h^2 \simeq 
 \mathcal{A}_{GW} \left(\frac{k}{10^4\text{Mpc}^{-1}}\right)^{n_k} \left(\frac{M_{\rm PBH}}{10^4 \text{g}}\right)^{n_M} \left(\frac{\Omega_{\rm PBH,f}}{10^{-10}}\right)^{n_{\Omega}} \left(\frac{\bar{\tau}_{\rm NL}{\cal P}_{\cal R}}{10^{-12}}\right)^{n_{\tau}} \,,
\end{align}
\end{widetext}
where the dimensionless amplitude $\mathcal{A}_{GW}$ and the power law indices $n_X$ ($X=k,M,\Omega,\tau$) are provided in Table~\ref{table:parameters}.

As one may see from the right panel of \Fig{fig:P_Phi_Omega_GW} we are met with a distinctive bi-peaked PBH induced GW signal with the high-frequency peak being related to the Gaussian component of the PBH matter spectrum and the low-frequency one associated to the non-Gaussian clustering term of $\mathcal{P}_{\delta_\mathrm{PBH}}(k)$. A future detection of such a bi-peaked GW signal will potentially act as a smoking gun of the presence of primordial non-Gaussianities on small scales. We need to note however that it is possible for two different set of parameters ($M_\mathrm{PBH}$, $\Omega_\mathrm{PBH,f}$,$\Bar{\tau}_\mathrm{NL}$) that the Gaussian and the non-Gaussian GW peaks coincide leading to a potential degeneracy between them. Nevertheless, since the spectral shapes around each GW peak are different [See \Eq{eq:OmegaGWall}], this will make them distinguishable at the observational level.

\begin{table}[htbp]
\caption{Values of dimensionless parameters introduced in the parameterization \Eq{eq:OmegaGWall}. 
$k_\mathrm{L}$ represents the scale below which the large $u$ and $v$ contributions dominate over the resonant contribution, thus exhibiting a different scaling behavior. }\label{table:parameters}
\begin{tabular}{cccccc}
\toprule
 $k$'s region & $\mathcal{A}_{GW}$ & $\quad n_k \quad$ & $\quad n_M\quad$ & $\quad n_{\Omega}\quad$ & $\; n_{\tau}\;$ \\
\midrule
$(2k_{\text{evap}} , 2k_{\rm{L}})$ & $1 \times 10^{-36}$ & $1$ & $71/18$ & $22/9$ & $2$ \\
$(2k_{\rm{L}} , 2k_{\rm d})$ & $2\times 10^{-72}$ & $17/3$ & $17/2$ & $0$ & $2$ \\
$(2k_{\rm d} , 2k_{\rm c})$ & $1\times 10^{-9}\,\,$ & $-7/3$ & $11/6$ & $16/3$ & $2$ \\
$(2k_{\rm c} , 2k_{\text{UV}})$ & $3\times 10^{-80}$ & $11/3$ & $41/6$ & $16/3$ & $0$ \\
\bottomrule
\end{tabular}
\end{table}

{\it{Constraining non-Gaussianities}}--
Remarkably, the aforementioned GW signal can be used as a novel probe to constrain NGs. In particular, one can impose constraints on $\bar{\tau}_\mathrm{NL}\mathcal{P}_\mathcal{R}(k)$ by accounting for theoretical and observational bounds on the GW amplitude. We stress that the non-Gaussian induced GWs studied in this Letter are produced after PBH evaporation and before BBN, hence they act as an extra relativistic component, being tightly constrained by BBN and CMB observations through the effective number of extra neutrino species $\Delta N_\mathrm{eff}<0.3$~\cite{Planck:2018vyg}. 
This upper bound constraint on $\Delta N_\mathrm{eff}$ can be translated to an upper bound on the GW amplitude which reads as $\Omega_\mathrm{GW,0}h^2\lesssim  10^{-6}$~\cite{Smith:2006nka,Caprini:2018mtu}, shown with the horizontal grey region in the right panel of \Fig{fig:P_Phi_Omega_GW}.  

Consequently, to avoid such a GW constraint from BBN one should require the GW amplitude at the two peaks, approximately located at $2k_{\rm d}$ and $2c_sk_{\rm UV}$, to be below $ 10^{-6}$. 
More specifically, the upper bound on the GW amplitude at $2 c_s k_\mathrm{UV}$ leads to an upper bound on the initial PBH abundance $\Omega_\mathrm{PBH,f}$~\cite{Domenech:2020ssp},
\beq\label{eq:Omega_f_upper}
\Omega_\mathrm{PBH,f}\lesssim 10^{-6}\left(\frac{M_\mathrm{PBH}}{10^4\mathrm{g}}\right)^{-17/24}\,.
\eeq
On the other hand, the upper bound on the GW amplitude at $2k_\mathrm{d}$ is translated to an upper bound on $\bar{\tau}_\mathrm{NL}\mathcal{P}_{\cal R}$, 
\beq\label{eq:tau_NL_upper_BBN}
\bar{\tau}_\mathrm{NL}\mathcal{P}_{\cal R}\big|_{k=2k_{\rm d}}\lesssim 4\times 10^{-20}\Omega_{\rm PBH,f}^{-17/9}\left(\frac{M_\mathrm{PBH}}{10^4\mathrm{g}}\right)^{-17/9}\,.
\eeq
In addition, the fact that the PBH isocurvature induced curvature power spectrum $\mathcal{P}^\mathrm{PBH}_\mathcal{R}(k)$, related to $\mathcal{P}_\Phi(k)$ as $\mathcal{P}^\mathrm{PBH}_\mathcal{R}(k) = \frac{9}{4}\mathcal{P}_\Phi(k)$, should be smaller than the inflationary curvature power spectrum $\mathcal{P}_{\cal R}(k)$, i.e $\mathcal{P}^\mathrm{PBH}_\mathcal{R}(k)<\mathcal{P}_{\cal R}(k)<10^{-2}$~\cite{Kalaja:2019uju,Gow:2020bzo}, leads to an upper bound on $\bar{\tau}_\mathrm{NL}\mathcal{P}_{\cal R}$, being recast as $\bar{\tau}_\mathrm{NL}\mathcal{P}_{\cal R}\big|_{k=2k_{\rm d}}\lesssim 2\times 10^{-3}$.

\begin{figure*}[h!]
\centering
\includegraphics[width=0.85\textwidth]{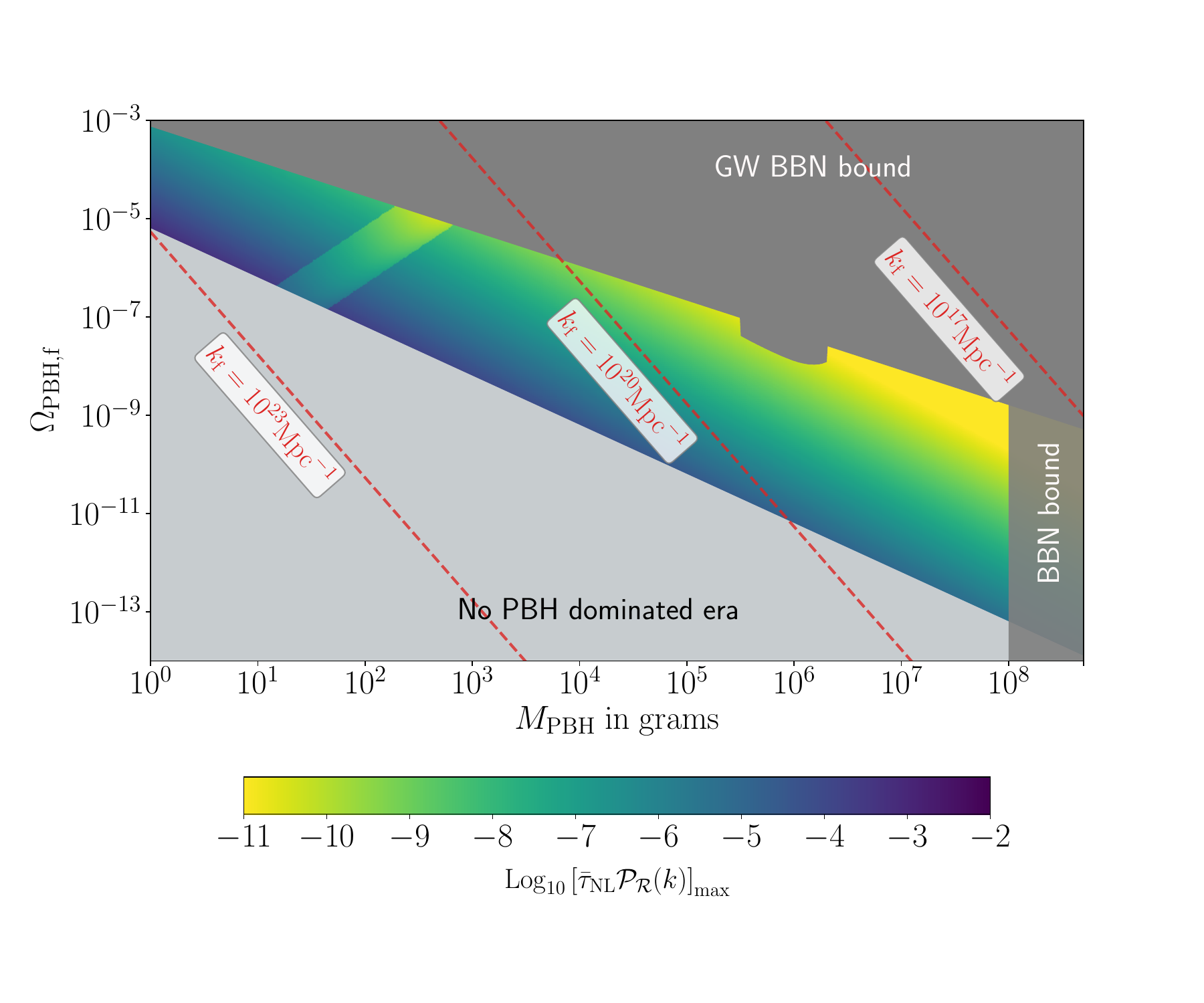}
\caption{\it{ The upper bound on $\bar{\tau}_\mathrm{NL}\mathcal{P}_{\cal R}(k)$ (color bar axis) as a function of the PBH mass $M_\mathrm{PBH}$ ($x$-axis) and the initial PBH abundance at formation $\Omega_\mathrm{PBH,f}$ ($y$-axis) accounting as well for the lower and upper bounds of $\Omega_\mathrm{PBH,f}$. }}
\label{fig:tau_NL_constraints}
\end{figure*}

In \Fig{fig:tau_NL_constraints} we show the upper bound constraint on $\bar{\tau}_\mathrm{NL}\mathcal{P}_{\cal R}(k)$ (color bar axis) as a function of the PBH mass $M_\mathrm{PBH}$ ($x$-axis) and the initial PBH abundance at formation $\Omega_\mathrm{PBH,f}$ ($y$-axis). The BBN bound on $\Omega_\mathrm{PBH,f}$, \Eq{eq:Omega_f_upper}, excludes the upper right dark gray region.
The quadrilateral pattern on the left side of the color map and the corresponding gap around $M_{\text{PBH}} \sim 10^6 \rm{g}$ arise from the constraints imposed by the LIGO-VIRGO-KAGRA (LVK) experiment  while the vertical grey region in the mass range $[10^8\mathrm{g},10^9\mathrm{g}]$ is excluded from the effect of Hawking evaporation of PBHs in this mass range on the abundances of light elements produced during BBN~\cite{Boccia:2024nly}. We also require that PBHs dominate before they evaporate. This leads to the lower bound on the $\Omega_\mathrm{PBH,f}$,
$\Omega_\mathrm{PBH,f}>6\times 10^{-10}\frac{10^4\mathrm{g}}{M_\mathrm{PBH}}$~\cite{Domenech:2020ssp}, which excludes the lower left light grey region.


{\it{Discussion}} -- 
In this Letter, we propose a novel probe of primordial NG, specifically targeting the non-linearity parameter $\tau_{\rm NL}$. This probe is powerful for the previously unconstrained ultra-light-mass window for PBHs, as primordial NG on small scales is sensitively captured by the spatial clustering of PBHs. Remarkably, we find a distinctive bi-peaked GW spectral shape, with the amplitude of the secondary GW peak at low frequencies being directly related to $\tau_{\rm NL}$. Notably, by accounting for upper bounds on the GW amplitude derived from existing theoretical and observational constraints, we establish for the first time to the best of our knowledge a joint constraint on the correlations between the effective non-linearity parameter $\bar{\tau}_{\rm NL}$ and the mass and abundance of ultra-light PBHs, which can effectively yield an upper bound on primordial NGs.

The proposed probe has several immediate cosmological implications. At the observational level, there is a promising chance for the GW signal induced by our mechanism to fall into the frequency detection bands of forthcoming GW experiments, pointing towards the potential detectability of the aforementioned distinctive GW signal in the near future.
At the theoretical level, one can test via the above GW portal any feasible inflation models that give rise to ultra-light PBHs with scale-dependent NGs. 
Additionally, it is worth mentioning that the novel NG probe introduced here can in principle be extended to any phenomenological scenario involving compact objects with a monochromatic or extended mass function~\cite{Papanikolaou:2022chm}, coupled with a transient eMD epoch. 
Those aforementioned subjects deserve to be studied in detail, and we leave related explorations to future work.

Finally, we need to highlight that future measurements and/or upper bound constraints on $\Delta N_\mathrm{eff}$ at the percent level in the coming years will set tighter constraints on local-type primordial non-Gaussianities on small scales through our PBH induced GW portal whereas forecasts on local-type non-Gaussianities from DESI~\cite{DESI:2023duv,Rezaie:2023lvi} and CMB-S4~\cite{Sohn:2019rlq} on larger scales (namely the LSS and CMB ones) will act as a complementary to our GW portal non-Gaussianity probe.

{\bf{Acknowledgments}}--We thank Guillem Dom\`enech, Keisuke Inomata, Takahiro Terada and Gabriele Franciolini for useful correspondence. This work is supported in part by National Key R\&D Program of China (2021YFC2203100), by NSFC (12261131497), by CAS (JCTD-2022-20), by 111 Project (B23042), by Fundamental Research Funds for Central Universities, by USTC Fellowships for International Cooperation, and by USTC Research Funds of the Double First-Class Initiative. It is also supported in part by the JSPS KAKENHI grant No. 20H05853.
TP and ENS acknowledge the contribution of the LISA Cosmology Working Group and the COST Action
CA21136 ``Addressing observational tensions in cosmology with systematics and 
fundamental physics (CosmoVerse)''. TP acknowledges also the support of INFN Sezione di Napoli \textit{iniziativa specifica} QGSKY as well as financial support from the Foundation for Education and European Culture in Greece.

\bibliography{PBH}

\end{document}